\documentstyle[12pt,aaspp4]{article} 
\def\tempest%
{\begin{array}{ccc} 
1 & 1 & 1 \\ 
1 & 1 & 1 \\ 
4 & 3 & 8 
\end{array}}

\begin{document}

\title{Complete Parallax and Proper Motion Solutions For Halo Binary-Lens 
Microlensing Events} 
\author 
{Andrew Gould\altaffilmark{1}\altaffiltext{1}{Alfred P.\ Sloan Foundation 
Fellow} and Nikolay Andronov} 
\affil{Ohio State University, Department of Astronomy, Columbus, OH 43210} 
\affil{E-mail: gould, andronov@astronomy.ohio-state.edu} 
\begin{abstract} 
A major problem in the interpretation of microlensing events is that the only 
measured quantity, the Einstein time scale $t_{\rm E}$, is a degenerate 
combination of the three quantities one would like to know, the mass, distance,
and speed of the lens. This degeneracy can be partly broken by measuring 
either a 
``parallax'' or a ``proper motion'' and completely broken by measuring both. 
Proper motions can easily be measured for caustic-crossing binary-lens events. 
Here we examine the possibility (first discussed by Hardy \& Walker) that one
could also measure a parallax for some of these 
events by comparing the light curves of the caustic crossing as seen from 
two observatories on Earth.  We derive analytic expressions for the 
signal-to-noise ratio of the parallax measurement in terms of the 
characteristics of the source and the geometry of the event.  
For Galactic halo binary lenses seen toward the 
LMC, the light curve is delayed from one continent to another 
by a seemingly minuscule 15 seconds (compared 
to $t_{\rm E}\sim 40\,$days). However, this is sufficient to cause a 
difference in magnification of order $10\%$. To actually extract complete 
parallax information (as opposed to merely detecting the effect) requires 
observations from three non-collinear observatories. Parallaxes cannot
be measured for binary lenses in the LMC but they can be measured for
Galactic halo binary lenses seen toward M31.  Robust measurements are possible
for disk binary lenses seen toward the Galactic bulge, but are difficult for
bulge binary lenses.

\end{abstract} 

\section{Introduction} 

One of the major problems in the interpretation of microlensing 
observations is that for most events the only physically relevant 
parameter extracted from the light curve is the Einstein time scale, 
$t_{\rm E}$, 
which is a complicated combination of three quantities that one would like 
to know individually; the mass of the lens, the distance to the lens, and 
the transverse speed of the lens relative to the observer-source line of 
sight. For example, more than a dozen candidate events have been found 
toward the Large Magellanic Cloud (LMC) (Alcock et al.\ 1997b; Aubourg et al.\ 
1993), but it is still not known if these are predominantly due to a new 
population of objects which comprise half or more of the mass of the halo, or 
if they are a previously unrecognized stellar structure either in the LMC 
itself (Sahu 1994; Wu 1994) or along the line of sight toward 
the LMC (Zhao 1998; 
Zaritsky \& Lin 1997; Evans et al.\ 1998). Similarly, several hundred events 
have been discovered toward the Galactic bulge (Udalski et al.\ 1994; 
Alcock et al.\ 1997a), and these could potentially be very useful to 
address questions of Galactic structure (Zhao, Rich, \& Spergel 1996) and 
the stellar mass function (Zhao, Spergel, \& Rich 1995; Han \& Gould 1996; 
Gould 1996a). However, the three-fold degeneracy among mass, distance, and 
speed makes such an analysis extremely difficult and subject to distortions 
from unknown systematic effects. 

A number of ideas have been advanced to partially or totally break 
this three-fold degeneracy. Gould (1992) showed that for sufficiently 
long events ($t_{\rm E}\ga {\rm yr}/2\pi$), the reflex motion of the Earth 
induces a distortion of the light curve which yields a ``parallax'', 
essentially the (two-components of the) transverse velocity of the lens 
projected onto the plane of the observer. Several parallaxes have now 
been measured toward the bulge (Alcock et al.\ 1995; Bennett 1997), and 
important information has been extracted from the lack of a parallax detection 
in a long event seen toward the Small Magellanic Cloud (SMC) 
(Palanque-Delabrouille et al.\ 1998; Afonso et al.\ 1998). Parallaxes 
cannot ordinarily be measured in this way for short events because the 
Earth does not change its velocity enough during the course of the event to 
produce a sufficient distortion of the light curve. However, even for 
short events one can sometimes gain some information from the ``parallax 
asymmetries'' induced in the light curve (Gould, Miralda-Escud\'e, \& Bahcall 
1994) and this information, while substantially less useful than a full 
parallax, could nevertheless be important in some applications 
(Buchalter \& Kamionkowski 1997; Gould 1998).

Parallaxes could be routinely measured for a large fraction of events 
by launching a satellite into solar orbit (Refsdal 1966; 
Gould 1994b; Gould 1995a; Boutreux \& Gould 1996; Gaudi \& Gould 1997; 
Markovi\'c 1998). Because 
the Einstein ring is usually of order a few AU, the light curve of the event 
is substantially different as seen from the Earth and the satellite. This 
enables one to determine essentially the time it takes for the event to 
get from one to the other and so (since the Earth-satellite distance is known) 
determine the projected transverse velocity. Unfortunately, no dedicated 
parallax satellite is currently planned. However, it will be possible to 
combine observations by the {\it Space Infrared Telescope Facility} with 
intensive ground-based observation to obtain parallaxes for some events 
(Gould 1999). 

	Hardy \& Walker (1995)
showed that it is possible to obtain some parallax information for binary-lens
events by comparing the light curves of the caustic crossing as seen from
two observatories on Earth.  The shorter baseline (relative to a satellite)
is compensated by the rapid change in the magnification.  They also showed
that by comparing observations from three non-collinear observatories, 
one could measure the full parallax.

A complementary type of additional information can be obtained from 
measurements of the ``proper motion'' of the lens relative to the 
observer-source line of sight. Because $t_{\rm E}$ is known, 
measuring the proper motion, $\mu$, is equivalent to measuring the angular size
of the Einstein ring, $\theta_{\rm E}=\mu t_{\rm E}$. Numerous ideas have 
been advanced 
to measure this quantity. If a single-lens transits the face of the source, 
the light curve will deviate from the point-source approximation, yielding 
the source transit time and so (since the angular size of the source is 
approximately known) the proper motion (Gould 1994a; Nemiroff \& Wickramasinghe
1994; Witt \& Mao 1994). The probability of such transits is low because 
the source size is much smaller than $\theta_{\rm E}$. However, Alcock et 
al.\ (1997c) have measured this effect for one bulge lensing event, and 
several other measurements have also been made but not yet published. Transits 
are extremely rare toward the LMC and SMC, in part because the source angular 
radii are smaller and in part because there are many fewer events. 
To date no single-lens transit events have been observed toward the LMC or SMC.
On the other hand, for binary-lens events, 
the proper motion can be determined whenever the source crosses the caustic 
(region of formally infinite magnification) by dividing the source size by 
the measured caustic-crossing time. Caustic-crossing binary-lens events 
comprise of order 5\% of all microlensing events, so this is a potentially 
very effective method. Measurement of the crossing time requires much more 
detailed coverage of the light curve than is available from the roughly 
nightly observations used to find microlensing events. However, three 
groups now intensively monitor ongoing events, 
GMAN (Alcock et al.\ 1997c), PLANET (Albrow et al.\ 1998), and MPS 
(Rhie et al.\ 1999). In particular, one of the only two events seen toward 
the SMC was a binary lens, 
and the measurement of its proper motion demonstrated 
that it is almost certainly in the SMC itself and not a halo lens 
(Afonso et al.\ 1998; Albrow et al.\ 1999; Alcock et al.\ 1999; Udalski et al.\
1998; Rhie et al.\ 1999). There are several other suggestions for measuring 
proper motions including a spectroscopic method (Maoz \& Gould 1994) and 
a binary-source method (Han \& Gould 1997). 

If both the parallax and the proper motion were measured, then there 
would be three measured quantities (including $t_{\rm E}$) and three 
unknowns (mass, distance, and speed) so a complete solution would be 
possible (Gould 1992, 1995b). To date, four practical ideas have 
been devised to obtain complete solutions, and each is applicable to only 
a relative handful of events. 

First, if a dedicated parallax satellite were launched, 
then parallaxes would be measured for most events, and the relatively small 
fraction for which proper motions could be obtained would then have complete 
solutions. Unfortunately, as mentioned above, no such mission is currently 
planned. 

Second, the proposed {\it Space Interferometry Mission} ({\it SIM}) could 
measure 
both the proper motion and the parallax of some events (Boden, Shao, \& 
Van Buren 1998). During microlensing 
events, the centroid of the two images is typically deflected by several 
tens of $\mu$as relative to the source position. The pattern of deviation is 
an ellipse whose size yields $\theta_{\rm E}$. Since {\it SIM} has an 
astrometric accuracy of $4\,\mu$as, it can measure this deviation quite well. 
The reflex motion of the Earth produces an additional deviation that is 
superimposed on this ellipse, and by measuring this deviation one can determine
the parallax. However, very few lensing events toward the Magellanic Clouds 
have sources brighter than $V=20$ (the {\it SIM} limit). Near the magnitude 
limit, very long integration times are required to reach the nominal precision 
of $4\,\mu$as, implying that it will be possible to obtain complete solutions 
for only a handful of events. Toward the bulge, the situation is much more 
favorable because there are many events with bright ($V\la 16$) sources. 
Nevertheless, since competition for {\it SIM} time will be severe, the 
total number of complete determinations will probably not be large. 

Third, complete solutions are possible for a significant fraction 
of extreme microlensing events (EMEs) (Gould 1997). By definition, 
EMEs have peak magnifications $A_{\rm max}\ga 200$. This high magnification 
permits measurement of the parallax by observing the event from two different 
locations from Earth. Recall that the reason for launching a parallax 
satellite was to get an observatory far enough away (in units of the Einstein 
ring) so that the event would appear significantly different. Two continents 
on Earth are separated by only $\sim 3\times 10^{-5}$ AU, so 
the fractional difference in magnification as seen 
from the two locations would ordinarily be of this order. 
However, Holz \& Wald (1996) pointed out that 
photon statistics alone do not necessarily prevent the detection of such 
a small effect. Moreover, the fractional difference is increased by 
approximately the magnification, so for EMEs the fractional difference in 
magnification between two observatories can be of order 1\%. Because 
the magnification is so high, there is a high probability that the lens will 
transit (or nearly transit) the source which would permit measurement of the 
proper motion. 
Gould (1997) estimated that of order 30 EMEs occur per year toward the Galactic
bulge, and that complete solutions could be obtained for a large fraction of 
them by follow-up observations. However, finding these EMEs would require 
a pixel-lensing (Gould 1996b) search of the entire bulge in real time. So 
far there are no plans to organize such a search. 

	Fourth, complete solutions can be obtained for some 
caustic-crossing binary events (Hardy \& Walker 1995).
The method is closely related to the EME method: caustic-crossing 
binary-lens events are automatically ``extreme magnification events'' since 
the caustic has formally infinite magnification. Actually, the peak 
magnification is suppressed by the finite size of the source, just as it is 
when a point lens transits a finite source. The peak magnifications 
generally do not get as high as for EMEs because the magnification scales 
as the inverse square root of the distance from the caustic for binary-lens 
caustics and scales
inversely for point lenses. Nevertheless, from the standpoint 
of measuring the parallax, what is important is not the magnification per 
se, but the logarithmic
rate of change of the magnification with position in the Einstein 
ring. Basically, the magnification goes from peak to nearly zero as the lens 
crosses the radius of 
the source. If the angular radius of the source is small, 
then the logarithmic magnification gradient can be very large. 

However, the size of the parallax effect does not depend only on the (known)
distance between the observatories: it depends on the difference 
in the distance between the source and the caustic as seen from the two 
observatories. This difference is equal to the distance between the two 
observatories {\it times} the cosine of the angle between the normal to 
the caustic and the line connecting the observatories. Hence, one cannot 
measure the parallax using two observatories alone: one can only obtain a 
lower limit. By adding a third observatory (not collinear with the other two), 
this degeneracy can be broken and the parallax measured. 
Since it is always possible to measure the proper motion of a 
caustic-crossing binary, those events with parallax measurements can 
be solved completely. 

	Here we investigate this method more closely.  In \S\ 2, we identify
the four quantities (in addition to the source distance) that must be 
measured to obtain a complete solution and
give explicit formulae for the mass, distance, and speed of the lens in
terms of these observables.  In \S\ 3, we present analytic expressions for
the difference in magnification as seen from two observatories for the case
when the source is close to but not yet crossing the caustic, and show that
the measurable quantity is a degenerate combination of the parameter one
would like to know and the secant of an unknown angle.  In \S\ 4 and \S\
5, we show that this degeneracy can be broken either by making observations
from a third non-collinear location or by observing two caustic crossings,
each from two observatories.  We indicate, however, that the latter is 
generally impractical.  In \S\ 6, we derive expressions for the
magnification and its derivative during a caustic crossing.  In \S\ 7, we 
show that the quantity most directly determined from a caustic-crossing
parallax measurement is $\tilde r_*$, essentially the physical size of the
source projected through the lens onto the observer plane.  We also show that
all three of the other measurable quantities identified in \S\ 2 depend 
on a correct modeling of the light curve as a whole, not just the caustic
crossing.  In \S\ 8, we derive analytic expressions for the signal-to-noise
ratio (S/N) in terms of the geometry of the event and the characteristics of 
the source and telescopes.  In particular, for sources above the sky, the $S/N$
depends only on the surface brightness of the source and not on its radius.
In addition, $S/N\propto D^{-1}$ where $D\equiv D_{os}D_{ol}/D_{ls}$ and
$D_{os}$, $D_{ol}$,  and $D_{ls}$ are the distances between the observer, 
source, and lens.  These facts allow us to classify all possible events
and to determine generally which have observable parallaxes.

We show that using 1 m class telescopes, it is possible to measure 
parallaxes for halo lenses seen toward the LMC and SMC, but it is not possible 
for LMC lenses and SMC lenses. One way to think about the difference is that 
the transverse velocity of a halo lens projected onto the Earth is about 
$275\,{\rm km}\,{\rm s^{-1}}$, so it takes about 15 seconds to sweep from one 
continent to another 4000 km away. However, the projected velocity of an LMC 
lens is about $1500\,{\rm km}\,{\rm s^{-1}}$, so the lens moves across the 
ocean in 3 seconds. The magnification simply does not change enough in 
3 seconds to detect the difference. 

We find that robust measurements can be made for foreground disk lenses 
seen toward the Galactic bulge, but only marginal detections should be expected
for bulge lenses. It should also be possible to measure the parallaxes of
relatively nearby ($\la 20\,$kpc) Galactic binary lenses detected toward M31,
but not for those that are substantially more distant.  
Two groups are currently searching for lensing events toward M31 
(Tomaney \& Crotts 1996; Ansari et al.\ 1997). While no binaries 
(and indeed no confirmed lensing events) have yet been detected, 
these experiments are just now gearing up to become major efforts. 

\section{Complete solutions}
In this paper we will show that is possible, at least in principle, to extract
two parameters from a binary-lens microlensing event, $t_*$ and $\tilde r_*$, 
\begin{equation}
t_* = \frac{D_{ol}\theta_*}{v},\qquad \tilde r_* = D\theta_*,\qquad
D\equiv\frac{D_{ol}D_{os}}{D_{ls}}. \label{shalala}   
\end{equation}
Here $\theta_*$ is the angular size of the source, $v$ is the transverse speed
of lens relative to the observer-source line of sight, and $D_{ol},D_{os},$
and $D_{ls}$ are the distances between the observer, lens, and source.
There are three other observable quantities:
\begin{equation}
\theta_*,\qquad t_{\rm E},\qquad D_{os}.
\end{equation}
The angular source size $\theta_* $ can be determined from its observed color
and magnitude, and the estimated extinction using the Planck law. The 
Einstein crossing time, 
\begin{equation}
t_{\rm E} = \frac{D_{ol}\theta_{\rm E}}{v},\qquad  \theta_{\rm E} \equiv 
\left(\frac{4 G M }{c^2 D}\right)^{1/2},
\end{equation}
can be determined from the overall fit to the light curve. Here 
$\theta_{\rm E}$ is the angular Einstein radius, and $M$ is the total mass of 
the binary lens. Finally, the distance $D_{os}$ is approximately equal to 
the mean distance of the source population (e.g.\ the LMC). 
 From these five parameters one can easily obtain the physically important 
quantities: 
\begin{equation}
D_{ol} = \left(\frac{\theta_*}{\tilde r_*}+\frac{1}{D_{os}}\right)^{-1},
\end{equation}
\begin{equation}
D_{ls}  = D_{os} \left(1+\frac{\tilde r_*}{\theta_* D_{os}}\right)^{-1},  
\end{equation}
\begin{equation}
v = \left(\frac{t_*}{\tilde r_*}+\frac{t_*}{D_{os}\theta_*}\right)^{-1},
\end{equation}
\begin{equation}
M = \frac{c^2}{4 G}\theta _* \tilde r_*\left(\frac{t_{\rm E}}{t_*}\right)^2.
\end{equation}

\section{Two observers}
 
Consider a microlensing event as observed from two different observatories. 
Let $d_{12}$ be the projected separation between the observatories, i.e. the 
physical distance between them {\it projected onto the plane of the sky.}
For a point source near a caustic the magnification as seen by each 
observatory is given by, 
\begin{equation}
A_i^0 = \alpha (\Delta u_i)^{-1/2}+\gamma,  \label{point}
\end{equation}
where $\Delta u_i$ is the angular separation between the source and the 
caustic in units of $\theta_{\rm E}$ as seen from each observatory (i = 1,2), 
and $\alpha$ and $\gamma$ are constants. In the range of interest,
\begin{equation}
d_{12} \ll \tilde r_{\rm E} ,\qquad \tilde r_{\rm E} \equiv D\theta_{\rm E},
\end{equation}
the caustic can be approximated as a straight line. Hence,
\begin{equation}
\Delta u_2 - \Delta u_1 = \frac{d_{12}}{\tilde r_{\rm E}}\cos\theta_{12}
\end{equation}
where $\theta_{12}$ is the angle between the projected separation of the 
observatories and the normal to the caustic. The ratio of magnification is 
therefore given by:
\begin{equation}
\frac{A_2^0}{A_1^0} = \frac{\alpha (\Delta u_1+ \frac{d_{12} \cos\theta_{12}}
{\tilde r_{\rm E}})^{-1/2}+\gamma}{\alpha (\Delta u_1)^{-1/2}+\gamma} \approx
1-\frac{1}{2}\frac{d_{12}}{\tilde r_{\rm E} \Delta u_1}
\cos\theta_{12}. \label{otnosh}
\end{equation}

\section{Three observers}
In equation $(\ref{otnosh})$ $\alpha$, $\gamma$, and $\Delta u_1$ are all 
known from the overall fit to the light curve. In addition, $d_{12}$ is 
known from terrestrial measurements.
Hence, the measurement of the flux ratio, $A_2^0/A^0_1$ yields only the
degenerate parameter combination, $\tilde r_{\rm E}\sec \theta_{12}$. To 
break this
degeneracy additional observations are required. In principle two
distinct types of additional observations could be used. First, one 
could observe the event from three observatories instead of two.
The only requirement is that the three observatories should not be collinear,
i.e. they must be vertices of a triangle. Now there are three unknowns
($\tilde r_{\rm E}$ , $\theta_{12}$, and $\theta_{13}$) but also 
three equations.
These are:\\
1) equation $(\ref{otnosh})$\\
2) a similar equation for the flux ratio $A_3^0/A_1^0$ (which yields 
$\tilde r_{\rm E}\sec \theta_{13}$), and \\
3) an additional expression that gives the relation between the angles, 
\begin{equation}
\theta_{12}+\theta_{13}=\theta_{213}. \label{ugli}
\end{equation}
Here $\theta_{213}$ is the known angle between the line connecting 
observers 1 and 2 and the line
connecting observers 1 and 3. Note that there is actually a sign ambiguity in 
equation ($\ref{ugli}$) for the relation among the angles: it could also be 
$|\theta_{12}-\theta_{13}|=\theta_{213}$. However, this ambiguity is easily 
resolved by
considering similar equations for $\theta_{132}$ and $\theta_{321}$.
Hence, with three observers, the degeneracy is broken, and $\tilde r_{\rm E}$ 
can be separately determined.

\section{Two caustic crossings} 
In principle, it is possible to break the degeneracy even in the case of 
two observers, provided that both observers monitor {\it two} caustic
crossings, $a$ and $b$. In this case, the angles at which the lens crosses
the caustic, $\phi_a$ and $\phi_b$, are known from the binary-lens solution
for the overall light curve. One can then measure $\tilde r_{\rm E}
\sec\theta_{12a}$
and $\tilde r_{\rm E}\sec\theta_{12b}$ and can break the degeneracy using the 
relation,
\begin{equation}
\theta_{12a}-\theta_{12b}=\phi_{a}-\phi_{b}.
\end{equation}
However, as a practical matter, there is little opportunity to make such 
measurements
because there is no warning of the first caustic, and hence there will
not be enough time to prepare for the measurements.

\section{The Magnification near the caustic}
The previous results concern the case where the source is small relative to 
its separation from the caustic. Let us now consider the case where the
source size and separation are comparable. The magnification of a point 
source continues to be
given by the equation $(\ref{point})$, but for a finite source we must 
integrate over the surface brightness of the source, 
\begin{equation}
A = \frac{\int_{0}^{r}\rho d \rho \int_{0}^{2 \pi}d \psi A^0 
[\Delta u(\rho , \psi)]J(\rho , \psi)}{\int_{0}^{r}\rho 
d \rho \int_{0}^{2 \pi}d \psi J(\rho , \psi)},
\end{equation}
where $\rho$ and $\psi$ are polar coordinates, $J(\rho,\psi)$ is the intensity
as a function of polar position,  $\Delta u(\rho,\psi) = \Delta u^0 +
(\rho/\theta_{\rm E})\cos\psi$,
and $\Delta u^0$ is the separation of the center of the star from the caustic.
For simplicity, we assume uniform surface brightness and find,
\begin{equation}
A(\eta) =  \alpha \left(\frac{\theta_{\rm E}}{\theta_*}\right)^{1/2}G(\eta) 
+\gamma,
\end{equation}
where,
\begin{equation}
G(\eta) = \frac{2}{\pi}\int_{\rm max(\eta,-1)}^{1}\left(\frac{1-x^2}{x-\eta}
\right)^{1/2}dx.
\end{equation} 
Here $\eta$ is the dimensionless separation between the source and the 
caustic, given by,
\begin{equation}
\eta = \frac{\Delta u^0 \theta_{\rm E}}{\theta_*}, \qquad (\eta<1).
\end{equation}
The derivative of $G$ is given by,
\begin{equation}
G^{\prime}(\eta)=\frac{2}{\pi}\int_{\max(\eta,-1)}^{1}\frac{x}
{[({x-\eta})({1-x^2})]^{1/2}}dx.
\end{equation}
The functions $G(\eta)$ and $G^{\prime}(\eta)$ are shown in Figure 1.

\section{Summary of Measurable Quantities}
Before continuing we pause to assess how the parameters 
$t_*$, $t_{\rm E}$, $\tilde r_*$, and $\theta_*$ depend on the 
observations, and to what degree their estimation depends
on the model of the binary lens which is derived from the 
full light curve.
Both $t_*$ and $t_{\rm E}$ can be derived from observations from a single
observatory, and both depend critically on a correct modeling of 
the binary lens. The shape of the light curve during a caustic crossing, 
$G(\eta)$, 
is shown in Figure 1. The caustic crossing time $\Delta t$ is defined
as the time necessary to move $\Delta \eta = 1$ in Figure 1, and 
is therefore quite robustly measured from the caustic-crossing data. However,
$t_* \equiv \Delta t\sin\phi$, where $\phi$ is the angle between 
the velocity of the lens relative to the source and the caustic. The 
determination of this
angle depends on the overall light curve, and good data over large parts 
of the light curve are necessary for an accurate measurement 
(e.g.\ Albrow et al.\ 1999). The determination of $t_{\rm E}$ also depends on 
the overall  light curve (e.g.\ Albrow et al.\ 1999).

By contrast, the determination of $\tilde r_*$ does not depend on the 
global light curve, but only on the caustic crossing. The measured quantity,
\begin{equation}
\frac{A(\eta_2)}{A(\eta_1)} = \frac{G(\eta_2)+\frac{\gamma}{\alpha}
\left(\frac{\theta_*}{\theta_{\rm E}}\right)^{1/2}}{G(\eta_1)+\frac{\gamma}
{\alpha}
\left(\frac{\theta_*}{\theta_{\rm E}}\right)^{1/2}}\approx 1+
\frac{G^{\prime}(\eta)}{G(\eta)}\Delta \eta,
\end{equation}
where,
\begin{equation}
\Delta \eta \equiv \eta_2 - \eta_1 =\frac{d_{12}\cos\theta_{12}}{\tilde r_*},
\end{equation}
depends only weakly on the binary-lens model parameters 
$\alpha$ and $\gamma$. Since $G^{\prime}(\eta)/G(\eta)$ is well determined
from the caustic-crossing data, and $d_{12}$ is known from terrestrial
measurements, the degenerate combination $\tilde r_* \sec\theta_{12}$ is well 
determined from the caustic-crossing measurements from two sites.
As discussed in \S\ 4, this degeneracy can be broken by observations from 
third site.
Finally, the color (say $V-I$) of the source can be determined from the 
approach to the second caustic crossing because the magnification is so high 
that blending 
plays very little role, and the star is not yet resolved by the caustic
so the magnified source has very nearly the same color as the intrinsic 
source. On the other hand determination of the intrinsic magnitude of the 
source is dependent on correct modeling of the decomposition of the observed 
flux into (magnified)
source and blend. The angular radius of the star, $\theta_*$, depends 
not only on the color and magnitude of the source, but also on the 
reddening. However this dependence is relatively weak (e.g.\ Albrow et al.\ 
1999). 
 
\section{Feasibility of making the measurement}

How practical is the measurement of $\tilde r_*$? There are two general 
requirements for the observations. First the relative difference of the two 
magnifications should be great enough to be recognized as a valid
result. We write this requirement in the form,
\begin{equation}
\frac{|A(\eta_2)-A(\eta_1)|}{A(\eta_1)}>p, \label{p1}
\end{equation}
where we suggest $p\sim 0.01$.
Since $A(\eta_2)-A(\eta_1)\cong\alpha (\theta_{\rm E}/\theta_*)^{1/2}
G^{\prime}(\eta)(\eta_2 - \eta_1)$,
equation $(\ref{p1})$ can be conveniently rewritten,
\begin{equation}
\frac{|G^{\prime}(\eta)\Delta \eta|}{G(\eta)+Z}>p, \qquad 
Z=\frac{\gamma}{\alpha}\left(\frac{\theta_*}
{\theta_{\rm E}}\right)^{1/2},\label{p11}
\end{equation}
where $\Delta\eta\equiv \eta_2 - \eta_1$. The left hand side of equation 
$(\ref{p11})$ reaches a maximum at $\eta=1$ (see Fig. 1). (Formally the maximum
is at $\eta=-1$, but in practice this peak is too short to be resolved). One 
may show analytically that $G^{\prime}(1)= -2^{1/2}$ and $G(1)=0$. Therefore, 
equation 
$(\ref{p11})$ may be rewritten,
\begin{equation}
\frac{|\Delta \eta|}{Z} = 2^{1/2}\frac{d_{12}|\cos\theta_{12}|/\tilde r_*}
{(\gamma/\alpha)(\theta_*/\theta_{\rm E})^{1/2}}>p,
\end{equation} 
or, 
\begin{equation}
0.15\alpha \left(\frac{\gamma}{5}\right)^{-1}\left(\frac{d_{12}}{3000\,
{\rm km}}\right)\left(\frac{D_{os}\theta_{\rm E}}{20\,{\rm AU}}\right)^{1/2}
\left(\frac{r_*}{1.5 r_{\odot}}\right)^{-3/2}\left(\frac{|\cos\theta_{12}|}
{0.7}\right)
\left(\frac{D_{ls}/D_{ol}}{5}\right)>p_1
\end{equation}

Hence, for halo lenses seen toward the LMC (for which the above 
normalizations of the parameters are ``typical''), the condition is 
relatively easily met. However, for LMC lenses ($D_{ls}/D_{ol}\sim 0.1$, 
$D_{os}\theta_{\rm E}\sim 3 AU$), 
the left hand side is smaller by a factor $\sim 130$, so parallax 
measurements would be extremely difficult. 

The second requirement is that the signal-to-noise ratio (S/N) be sufficiently 
high for robust detection. For an exposure time $t_{\rm exp}$, the 
S/N is given by, 
\begin{equation} 
{S\over N} 
= {\alpha (t_{\rm E}/t_*)^{1/2}|G'(\eta)\Delta\eta|F_*\Gamma t_{\rm exp} 
\over 
(2\{ [\alpha(t_{\rm E}/t_*)^{1/2}G(\eta) + \gamma]F_* + F_{\rm sk}\}\Gamma 
t_{\rm exp})^{1/2}},\label{eqn:sngen} 
\end{equation} 
where $F_*$ is the unmagnified flux of the source, $\Gamma$ is the rate 
of photon detection per unit flux, and $F_{\rm sk}$ is the flux from the sky 
within the aperture of the point spread function (PSF). 
For a typical observing system and a $V_*=20$ source, 
$F_* \Gamma = 25\,{\rm s}^{-1} ({\cal D}/\rm m)^2$, where ${\cal D}$ is the 
diameter of the mirror. Note that we have made use of the fact that 
$t_{\rm E}/t_* = \theta_{\rm E}/\theta_*$. We adopt $F_{\rm sk}$ equivalent 
to $V_{\rm sk}=20$ which corresponds to a sky brightness of 
$V=21.3\ \rm mag\ arcsec^{-2}$ and 
a $1''$ PSF. 

Equation (\ref{eqn:sngen}) can be rewritten, 
\begin{equation} 
{S\over N} = {\alpha|G'(\eta)|\over 
[\alpha(\theta_{\rm E}/\theta_*)^{1/2}G(\eta) + B]^{1/2}} 
\biggr[{ F_*\Gamma t_{\rm exp}\over 2}\biggl({t_{\rm E}\over t_*}\biggr) 
\biggr]^{1/2}\,{d_{12}|\cos\theta_{12}|\over\tilde r_*},\label{eqn:snone} 
\end{equation} 
where, 
\begin{equation} 
B= \gamma + {F_{\rm sk}\over F_*}.\label{eqn:bdef} 
\end{equation} 
Again, inspection of Figure (1) shows that the S/N will be maximized near the 
end of the caustic crossing, $\eta=1$, where $G'(\eta)=-2^{1/2}$ and 
$G(\eta)=0$. In this limit, the first term of equation (\ref{eqn:snone}) 
approaches $(2/B)^{1/2}$. Hence, there are two limits, depending on 
whether the magnified source {\it outside the caustic} is brighter or 
fainter than the sky integrated over the PSF aperture. 

If the sky dominates, the S/N is given by, 
\begin{eqnarray} 
{S\over N} = 13\,\alpha\, 10^{(V_{\rm sk} - V_*)/5} 
\biggl({S_*\over S_\odot}\biggr)^{1/2} 
\biggl({D_{os}\over 50\,\rm kpc}\biggr)^{-1} 
\biggl({t_{\rm E}\over 40\,\rm day}\biggr)^{1/2} 
\biggl({D_{ls}/D_{ol}\over 5}\biggr) \nonumber \\
\times\biggl({t_{\rm exp}\over t_*/10}\biggr)^{1/2} 
\biggl({d_{12}\over 3000\,\rm km}\biggr) 
\biggl({|\cos\theta_{12}|\over 0.7}\biggr) 
\biggl({{\cal D}\over 1\,\rm m}\biggr), 
\qquad({\rm sky}\ {\rm dominates}),\label{eqn:snskylim} 
\end{eqnarray} 
where $S_*$ is the (reddened) surface brightness of the star, $S_\odot$ 
is the surface brightness of the Sun, and where we have made use of the 
fact that $F_*/r_*^2 = S_*/D_{os}^2$. If the source dominates, then 
\begin{eqnarray} 
{S\over N} = 6\alpha \biggl({\gamma\over 5}\biggr)^{-1/2} 
\biggl({S_*\over S_\odot}\biggr)^{1/2} 
\biggl({D_{os}\over 50\,\rm kpc}\biggr)^{-1} 
\biggl({t_{\rm E}\over 40\,\rm day}\biggr)^{1/2} 
\biggl({D_{ls}/D_{ol}\over 5}\biggr) \nonumber \\
\times \biggl({t_{\rm exp}\over t_*/10}\biggr)^{1/2} 
\biggl({d_{12}\over 3000\,\rm km}\biggr) 
\biggl({|\cos\theta_{12}|\over 0.7}\biggr) 
\biggl({{\cal D}\over 1\,\rm m}\biggr), 
\qquad({\rm source}\ {\rm dominates}).\label{eqn:snsourcelim} 
\end{eqnarray} 
For simplicity, we have focused on the $S/N$ achieved during the last $0.1 t_*$
of the caustic crossing. The total $S/N$ could be improved by a factor $\sim 2$
by monitoring the entire crossing which, for a halo binary lens toward the 
LMC, lasts about $3t_*\sim 30$ minutes.
These results indicate that for halo binary lenses, parallax is 
measurable only for sources down to about two magnitudes fainter than the sky 
($V_*\la 22$), and then only for sources hotter than the Sun. Even for 
sources that are brighter than the sky (when magnified just outside the 
caustic), the source should be bluer than the Sun to obtain a good 
measurement of the parallax. 
In particular, red giant sources (no matter how bright) will have relatively 
low $S/N$. Equations (\ref{eqn:snskylim}) and (\ref{eqn:snsourcelim}) show 
that it 
is not possible (with modest-size telescopes) to measure parallaxes for 
LMC binary lenses (for which $D_{ls}/D_{ol}\la 0.1$).
 
To consider other lines of sight, first note that the factors
$(D_{os}/50\,{\rm kpc})^{-1}(D_{ls}/D_{os}/5)$ can be written more simply
as $(D/10\,{\rm kpc})^{-1}$.  See equation (\ref{shalala}).

For bulge events, the optimal sources are turnoff stars because
they are the bluest common bulge stars.  
Taking account of extinction
($A_V\sim 1.5$), we evaluate the parameter combination in equations 
(\ref{eqn:snskylim}) and (\ref{eqn:snsourcelim})
$(S_*/S_\odot)(D/10\,{\rm kpc})^{-1}\sim D_{ls}/D_{os}$.  Thus parallaxes
could be easily measured for foreground disk binary lenses
($D_{ol}\ga D_{ls}$),
and with some difficulty for bulge binaries $(D_{ls}/D_{ol}\sim 1/3)$.

Parallaxes might also be measured for Galactic halo binary lenses
observed toward M31.  In this case 
$(D/10\,{\rm kpc})^{-1}\simeq (D_{ol}/10\,{\rm kpc})^{-1}$.  Thus, especially
for blue main-sequence sources (with their high surface brightness)
reasonable S/N could be obtained out to $D_{ol}\sim 20\,\rm kpc$.

The requirement that the observations be done at night (which is 
usually taken for granted) imposes considerable additional constraints in
the present case.  First, parallax measurements are generally possible
only during the autumn and winter because widely separated observatories
are not usually in darkness at the same time near the summer solstice.
For the LMC and SMC in particular, the observations must be done in autumn
and winter because the only suitable (non-collinear) location for a third
observatory is Antarctica which is in daytime during the entire spring and
summer.  Note that for the LMC this is particularly awkward since the
LMC is under the pole in winter.  Autumn and winter are, of course, the
most favorable times to observe the bulge.  Moreover, there are numerous
northern observatories around the globe that can view the bulge, at least
for brief periods, and which could therefore serve as third 
non-collinear observatories.  M31 observations are most feasible in October
when it is up all night and when the nights are reasonably long.  Since
most northern observatories are at $\sim 30^\circ$ latitude, it will be
somewhat difficult to obtain a long north-south baseline for the third
non-collinear observatory.


{\bf Acknowledgements}: 
This work was supported in part by grant AST 97-27520 from the NSF.

\clearpage 
\begin{figure}
\caption[junk]{\label{fig:one}
Normalized light-curve profile $G(\eta)$ ({\it solid curve})
for a uniform source crossing a
binary-lens caustic.  The magnification is given by
$\alpha (\theta_{\rm E}/\theta_*)^{1/2}G(\eta) + \gamma$, 
where $\alpha$ and $\gamma$ are constants, $\theta_{\rm E}$ is the angular
Einstein radius, $\theta_*$ is the angular radius of the source,
and $\eta$ is the separation of
the source center from the caustic in units of $\theta_*$.  The
sign convention is such that $\eta>0$ if the source is outside the caustic.
The parallax effect (difference in magnifications as seen from two
observatories) is $\alpha (\theta_{\rm E}/\theta_*)^{1/2}G'(\eta)\Delta \eta$, 
where $G'$ ({\it bold curve}) is the derivative of $G$, 
$\Delta\eta = d_{12}\cos\theta_{12}/\tilde r_*$ is the difference in values
of $\eta$ between the two observatories, $d_{12}$ is the distance between the
observatories projected on the plane of the sky, $\theta_{12}$ is the angle
between the normal to the caustic and the line connecting the observatories,
$\tilde r_*=D\theta_*$ is the source 
radius projected onto the plane of the observer, $D=D_{os}D_{ls}/D_{ol}$,
and $D_{ol}$, $D_{ls}$, and $D_{os}$ are the distances between the observer,
lens, and source.  Near the end of the caustic crossing $(\eta=1)$, the 
figure shows that
$|G'(\eta)|\rightarrow 2^{1/2}$, while $G(\eta)\rightarrow 0$, so the
fractional difference in magnification reaches a maximum.
}
\end{figure}

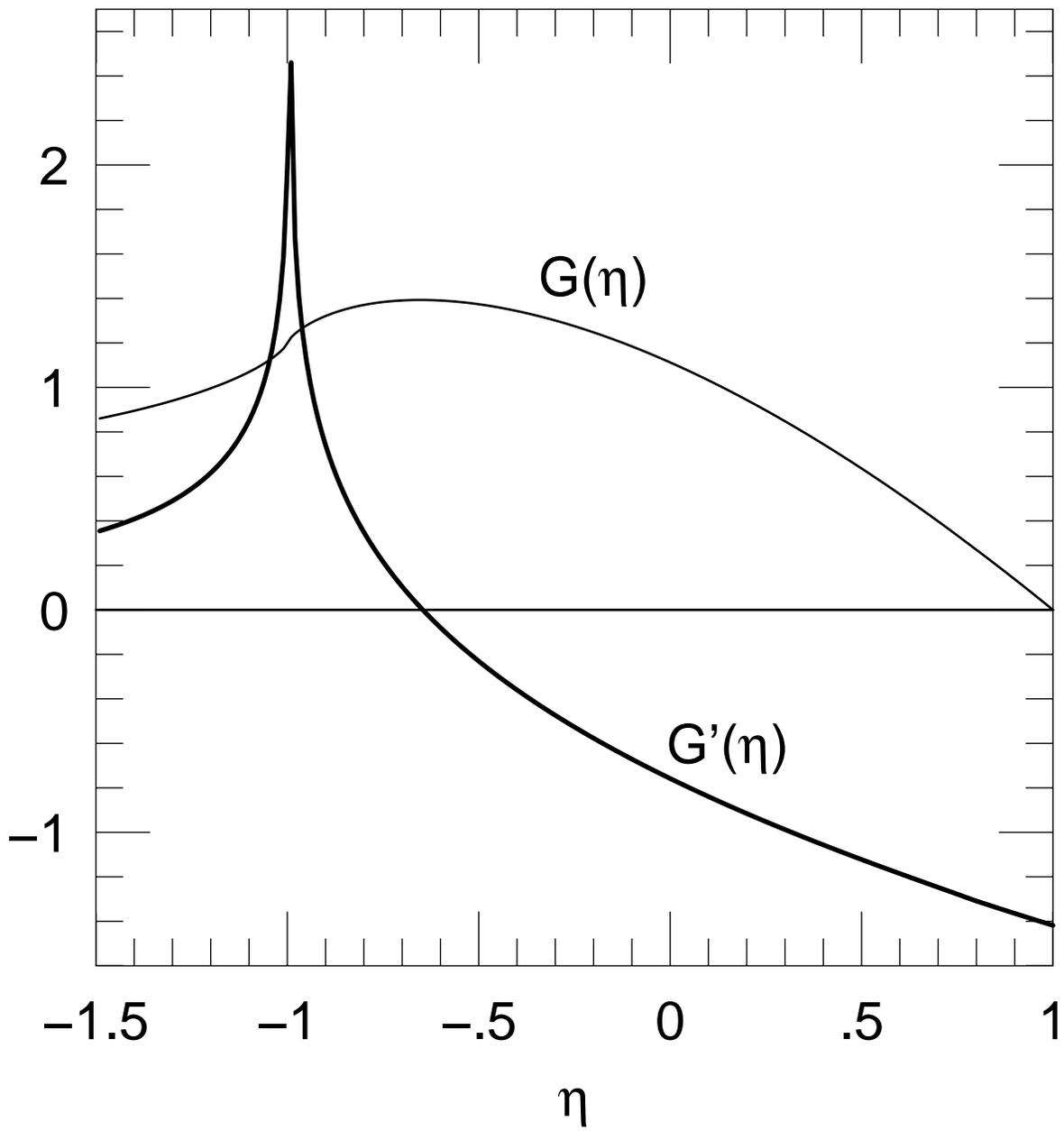
\end{document}